\documentclass[twocolumn,showpacs,amsmath,amssymb]{revtex4}
\usepackage{graphicx}

\begin{document}

\title{Multiscale Entanglement Renormalization Ansatz for Kondo Problem}
\author{Hiroaki Matsueda\footnote{matsueda@sendai-nct.ac.jp}}
\affiliation{Sendai National College of Technology, Sendai 989-3128, Japan}
\date{\today}
\begin{abstract}
We derive the multiscale entanglement renormalization ansatz (MERA) for the single impuity Kondo model. We find two types of hidden quantum entanglement: one comes from a finite-temperature effect on the geometry of the MERA network, and the other represents screening of the impurity by conduction electrons. As the latter starts to dominate the electronic state, the Kondo physics emerges. The present result is a simple and beautiful example of a holographic dual of a boundary conformal field theory.
\end{abstract}
\pacs{05.10.Cc, 11.25.Hf, 11.25.Tq, 04.70.Dy, 89.70.Cf}
\maketitle

The multiscale entanglement renormalization ansatz (MERA) offers the best variational wave function for quantum critical systems~\cite{Vidal}. The MERA is also recognized to be a discrete version of the anti-de Sitter space / conformal field theory (AdS/CFT) correspondence that has been a hot topic in superstring theory~\cite{Maldacena,Swingle}. Thus, deeper understanding of its functionalities similar to those of AdS/CFT and constructing efficient analytical or numerical methods for real applications to condensed matter physics are necessary.

In this paper, we apply MERA to the single impurity Kondo model~\cite{Kondo,Andrei}. The MERA network represents a holographic space where quantum entanglement propagates. By introducing a finite-temperature MERA network, we can clearly see the entanglement structure between the impurity and conduction electrons~\cite{Matsueda}. Then, the Kondo temperature $T_{K}$ can be obtained from the MERA network itself without assuming the specific Kondo interaction. Therefore, $T_{K}$ is a universal parameter determined from the global geometric structure of the holographic space. Furthermore, in connection with AdS/CFT, we argue that the Kondo physics is one of condensed-matter realization of black hole thermodynamics. Since the Kondo Hamiltonian is the most typical correlated electron system and its critical behaviors have been well known, some string theorists have recently attacked this problem by means of AdS/CFT~\cite{Harrison,Benincasa}. However, the theoretical treatments are in some sense far from condensed-matter terminologies. Our work translates them into much simpler MERA language.

We start with the impurity Kondo model in spatially one dimension which is a good example to study boundary critical phenomena and the local Fermi liquid theory. The Hamiltonian is given by
\begin{eqnarray}
H = -\sum_{i,j}t_{ij}c_{i}^{\dagger}c_{j} + J c_{1}^{\dagger}\vec{\sigma}c_{1}\cdot\vec{S},
\end{eqnarray}
where $c_{i}$ ($c_{i}^{\dagger}$) is spinor representation of an electron annihilation (creation) operator at site $i$, $\vec{S}$ is a local impurity with spin $S$, $t_{ij}$ is electron hopping, and $J$ is Kondo coupling.

\begin{figure}[htbp]
\begin{center}
\includegraphics[width=7.5cm]{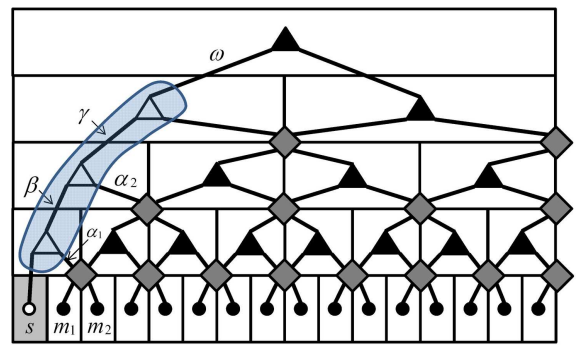}
\end{center}
\caption{The MERA network in the impurity Kondo model. Filled dots, diamonds, and filled triangles are the original sites in a quantum 1D critical system, disentangler tensors, and isometries (except for the top tensor), respectivery. An open circle represents the impurity site, and open triangles (shaded area) are boundary tensors. Note that the vertical direction denotes renormalization flow.}
\label{fig1}
\end{figure}

We consider a variational ansatz appropriate for the Kondo Hamiltonian. Because of the criticality, the ansatz has a MERA network with some modification near the impurity site. Figure~\ref{fig1} is the graphical representation of the MERA network. The impurity is located at the edge of half-infinite chain, but the impurity can interact with the conduction electrons at any site through the holographic branches. Thus, we believe that this configuration is a good starting point. The variational ansatz in the ground state is represented as
\begin{eqnarray}
\left|\Psi\right> &=& \sum_{s,\{m_{j}\}}C^{s,m_{1}m_{2}\cdots m_{L}}\left|s,m_{1}m_{2}\cdots m_{L}\right> ,
\end{eqnarray}
where $s$ is degree of freedom of the impurity, and $\{m_{j}\}$ are degrees of freedom of the conduction electrons. The coefficient $C^{s,m_{1}m_{2}\cdots m_{L}}$ can be decomposed into a product of three tensors
\begin{eqnarray}
C^{s,m_{1}m_{2}\cdots m_{L}} &=& \sum_{\{\alpha_{j}\},\alpha,\omega}T^{\omega}_{\alpha}B^{s,\omega}_{\alpha_{1}\alpha_{2}\cdots\alpha_{N}}\Phi_{\alpha_{1}\alpha_{2}\cdots\alpha_{N},\alpha}^{m_{1}m_{2}\cdots m_{L}} ,
\end{eqnarray}
where the number of layers is given by $N=\log_{2}L-1$, $T_{\alpha}^{\omega}$ is the top tensor, $\Phi_{\alpha_{1}\alpha_{2}\cdots\alpha_{N},\alpha}^{m_{1}m_{2}\cdots m_{L}}$ is the MERA network with active indices $\{\alpha_{j}\}$ ($j=1,2,...,N$) and $\alpha$, and the boundary tensor $B^{s,\omega}_{\alpha_{1}\alpha_{2}\cdots\alpha_{N}}$ (like the 'skin' of an onion $\Phi$) is defined by the matrix product state (MPS) form
\begin{eqnarray}
B^{s,\omega}_{\alpha_{1}\alpha_{2}\cdots\alpha_{N}} &=& \sum_{\beta,\gamma,...,\psi} W^{s\beta}_{\alpha_{1}}W^{\beta\gamma}_{\alpha_{2}}\cdots W^{\psi\omega}_{\alpha_{N}} , \label{boundary}
\end{eqnarray}
where $W$ is the isometry tensor. The indices $\{\alpha_{j}\}$ represent how the impurity interacts with the conduction electrons. Here we take $\alpha_{j}=1,2,...,M_{j}$. Note that the boundary tensor connects with the MERA network of the conduction electrons only through the disentangler tensors and the top tensor as shown in Fig.~\ref{fig1}. In addition, the boundary tensor does not contain disentanglers. In the continuous limit, the distance between the nearest neighbor isometries at the boundary is different from that in the bulk, since all of the isometries and the disentanglers are the same lattice points. This will lead to the difference between the bulk and the boundary scaling dimensions. Thus, our MERA network naturally describes the boundary CFT feature~\cite{Cardy,McAvity,Affleck,Pfeifer}.

Let us move to finite-temperature cases. In our recent paper associated with finite-temeperature MERA, it is possible to give a gravitational interpretation to the geometry of the holographic space~\cite{Matsueda}. As shown below, the MERA network is doubled, and two copies are combined with each other in IR region of the network. Then, we have found that the interface of the combined networks can be regarded as an event horizon. Let us imagine that this statement is also meaningful for the interface between $B^{s,\omega}_{\alpha_{1}\alpha_{2}\cdots\alpha_{N}}$ and $\Phi_{\alpha_{1}\alpha_{2}\cdots\alpha_{N},\alpha}^{m_{1}m_{2}\cdots m_{L}}$ (we suppose that $B^{s,\omega}_{\alpha_{1}\alpha_{2}\cdots\alpha_{N}}$ represents the impurity degree of freedom). This means that the impurity behaves as a black hole. Physically, this would be reasonable, since the screening of the impurity by the conduction electrons occurs, and some information associated with the spin degrees of freedom is lost due to their singlet formation. Since the singlet is the maximally entangled state, the maximally entanglement entropy is actually equivalent to the black hole entropy~
\cite{Bombelli,Susskind,Fiola,Hawking,Emparan,Solodukhin}.

\begin{figure}[htbp]
\begin{center}
\includegraphics[width=7.5cm]{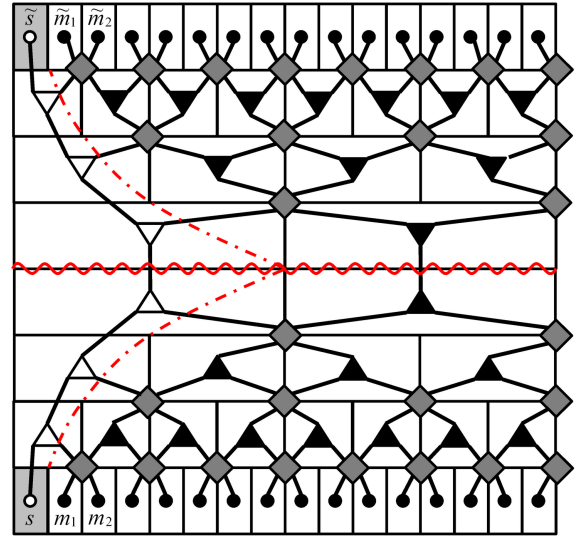}
\end{center}
\caption{Dual MERA network. A red wavy line represents an event horizon arizing from finite-temperature effect, while a dashed line represents a horizon associated with entanglement between the impurity and the conduction electrons.}
\label{fig2}
\end{figure}

\begin{figure}[htbp]
\begin{center}
\includegraphics[width=7.5cm]{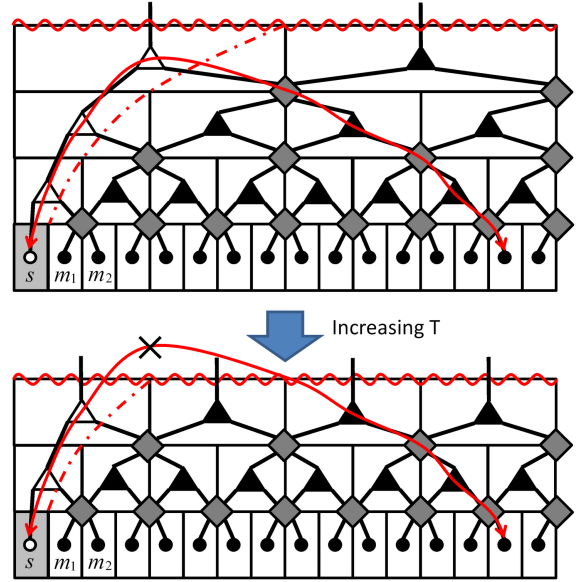}
\end{center}
\caption{Spin correlation in the holographic space. When the horizon coming from the impurity is terminated by the wavy horizon, the long range spin correlation is lost.}
\label{fig3}
\end{figure}

At finite temperatures, the wave function is described by using thermofield dynamics~\cite{Takahashi,Israel,Maldacena2,Cantcheff,Czech,Raamsdonk}. As already examined in the previous paper~\cite{Matsueda}, the thermal state is given by the MERA - tilde MERA combined network:
\begin{eqnarray}
\left|\Psi(\beta)\right> &=& \sum_{s,\{m_{j}\}}\sum_{\omega,\sigma}C_{\omega,\sigma}^{s,m_{1}m_{2}\cdots m_{L}}C_{\omega,\sigma}^{\tilde{s},\tilde{m}_{1}\tilde{m}_{2}\cdots\tilde{m}_{L}} \nonumber \\
&& \times \left|s,m_{1}m_{2}\cdots m_{L}\right>\left|\tilde{s},\tilde{m}_{1}\tilde{m}_{2}\cdots\tilde{m}_{L}\right> ,
\end{eqnarray}
and $C_{\omega,\sigma}^{s,m_{1}m_{2}\cdots m_{L}}$ is truncated MERA network~\cite{Javier,Javier2} defined by
\begin{eqnarray}
C_{\omega,\sigma}^{s,m_{1}m_{2}\cdots m_{L}}
&=& \sum_{\{\alpha_{j}\}}B^{s,\omega}_{\alpha_{1}\alpha_{2}\cdots\alpha_{N}}\Phi_{\sigma,\alpha_{1}\alpha_{2}\cdots\alpha_{N}}^{m_{1}m_{2}\cdots m_{L}} ,
\end{eqnarray}
where $N$ is the number of layers along the vertical direction. It is noted that the index $\sigma$ is put onto the MERA network for the conduction electrons.

Figure~\ref{fig2} is the graphical representation of the network. We find that the indices $\sigma$ and $\omega$ represent degrees of freedom of the event horizon arizing from finite temperature effect (we may neglect the effect of $\omega$ on the entanglement entropy for large $L$). On the other hand, the $\alpha$ parameters represent degrees of freedom for entanglement between the local impurity and the conduction electrons. Therefore, two types of different horizons appear. Strictly saying, the latter is artificial one, since the black hole should contribute to thermalization of the system that only comes from the former one as shown later.

When the latter dominates the electronic states, the impurity spin can be efficiently screened by the conduction electrons. This is holographic representation of the Kondo effect. Figure~\ref{fig3} shows the spin correlation in the holographic space. At zero temperature (Fig.~\ref{fig1}), the correlation is long-ranged. As temperature increases, some upper layers of the MERA network start to truncate. Then, the wavy horizon terminates the impurity horizon, and the long-range spin correlation is lost. The interesting thing is that we do not explicitely treat the Hamiltonian. Of course, our MERA network has information about the location of the impurity and the singlet formation among the impurity and the conduction electrons. However, the information is indirect. We think that even such loose conditions are enough for describing critical behavior of this system.

In the following, we determine the Kondo temperature $T_{K}$ from the MERA network. For this purpose, we calculate the black hole entropy by based on the holographic theory~\cite{Matsueda,Azeyanagi,Swingle2}. The entropy is given by the logarithm of the total degrees of freedom at the two horizons $\chi$ as
\begin{eqnarray}
S_{BH} &=& \ln\chi . \label{cond1}
\end{eqnarray}
Here, our original discrete network in Fig.~\ref{fig1} has the following metric in the continuous limit ($\eta=2$ in a case of binary MERA network):
\begin{eqnarray}
ds^{2}=\bigl(d\tau\ln\eta\bigr)^{2} + \bigl(\eta^{-\tau}dx\bigr)^{2} .
\end{eqnarray}
When we define the variable $z$ as
\begin{eqnarray}
z=\eta^{\tau}, \label{transform}
\end{eqnarray}
we obtain the standard notation of the AdS space
\begin{eqnarray}
ds^{2}=\frac{l^{2}}{z^{2}}\left(dz^{2}+dx^{2}\right).
\end{eqnarray}
In the present case, $\chi$ is represented as
\begin{eqnarray}
\chi = m^{A}\prod_{j=1}^{N}M_{j} ,
\end{eqnarray}
where $N(=\tau_{0})$ is the bond number at the interface between $B$ and $\Phi$ (or the layer number of the MERA network, since the bond number is $1$ per layer), and $A$ is the bond number at the MERA - tilde MERA interface. When the impurity is disentangled from the conduction electrons ($M_{j}=1$ for all $j$), the previous result is obtained~\cite{Matsueda}. Here, $m$ is the rank of the tensor at the latter interface. The area $A$ is given by
\begin{eqnarray}
A=\frac{L}{\eta^{N}} , \label{key}
\end{eqnarray}
and for simplitity we take $\eta=e$, although our figures are binary MERA networks. The paramter $m$ is the tensor dimension at the horizon. Then, we find
\begin{eqnarray}
S_{BH} = \frac{L}{e^{N}}\ln m + \sum_{j=1}^{N}\ln M_{j} . \label{entropytau}
\end{eqnarray}
By changing the variable $\tau_{0}$ to $z_{0}$ according to Eq.~(\ref{transform}), $S_{BH}$ is finally represented as
\begin{eqnarray}
S_{BH} &=& \frac{L}{e^{N}}\ln m + N\ln m + \sum_{j=1}^{N}\ln\left(\frac{M_{j}}{m}\right) \\
&=& S_{bulk} + S_{boundary}, \nonumber
\end{eqnarray}
where the bulk part $S_{bulk}$ and the boundary part $S_{boundary}$ are respectively defined by
\begin{eqnarray}
S_{bulk} &=& \ln\left(\frac{z_{0}}{2}e^{L/z_{0}}\right)\ln m , \label{entropyz} \\
S_{boundary} &=& \ln 2 \ln m + \sum_{j=1}^{N}\ln\left(\frac{M_{j}}{m}\right) . \label{boundary2}
\end{eqnarray}
(We may have some umbiguity for the determination of the boundary part due to the discretization of our spacetime.)

Equation~(\ref{entropyz}) should be consistent with the bulk entanglement entropy evaluated by the CFT living in the UV boundary of the MERA network~\cite{Azeyanagi,Swingle2}. The entropy is actually given by
\begin{eqnarray}
S_{EE} &=&\frac{c}{6}\ln\left(\frac{\beta}{\pi}\sinh\left(\frac{\pi L}{\beta}\right)\right) \nonumber \\
&\sim&\frac{c}{6}\ln\left(\frac{\beta}{2\pi}\exp\left(\frac{\pi L}{\beta}\right)\right) , \label{cond3}
\end{eqnarray}
for small $T$ with a finite temperature resolution $L^{-1}$~\cite{Holzhey,Calabrese}. Here, $c$ is the central charge, and the UV cutoff is taken to be unity. Identifying Eq.~(\ref{entropyz}) with Eq.~(\ref{cond3}), we find the temperature of this system as
\begin{eqnarray}
k_{B}T=\frac{1}{\pi z_{0}}=\frac{1}{\pi}e^{-\tau_{0}} . \label{temp}
\end{eqnarray}

Equation~(\ref{boundary2}) comes from the $g$ function of the boundary CFT~\cite{Affleck,Takayanagi}. It is clear that we have the following relations at the horizon when the three indices of the boundary isometry have the same dimensions:
\begin{eqnarray}
m = M_{N} = e^{c/6}.
\end{eqnarray}
Thus, the result indicates that the boundary entropy $\ln(M_{j}/m)$ at layer $j$ decreases under the boundary RG flow. This seems to be consistent with the conjectured $g$ theorem. Since the IR behavior of the Kondo model is described by the local Fermi liquid~\cite{Nozieres}, the effect of the impurity on the network is smeared out in the IR region.

The screening of the impurity spin becomes dominant, when the first term in Eq.~(\ref{entropytau}) is greater than the second term. The boundary is determined by solving the following equation for $N_{K}$
\begin{eqnarray}
\frac{L}{e^{N_{K}}}\ln m = \sum_{j=1}^{N_{K}}\ln M_{j} , \label{nk}
\end{eqnarray}
and the Kondo temperature is given by Eq.~(\ref{temp}) with use of $N_{K}(=\tau_{K})$:
\begin{eqnarray}
k_{B}T_{K}=\frac{1}{\pi z_{K}}=\frac{1}{\pi}e^{-\tau_{K}} . \label{kondo}
\end{eqnarray}
Let us compare this result with the well-known Kondo temperature $k_{B}T_{K}\sim \exp\left(-1/J\rho\right)$ where $\rho$ is the density of states. It is hard to solve Eq.~(\ref{nk}), but it's easy to find that $\tau_{K}$ increases with $L$. Thus, $\tau_{K}$ increases with decreasing the impurity density and $\rho$. Actually, this feature is not contradict to the exact $T_{K}$.

\begin{figure}[htbp]
\begin{center}
\includegraphics[width=7.5cm]{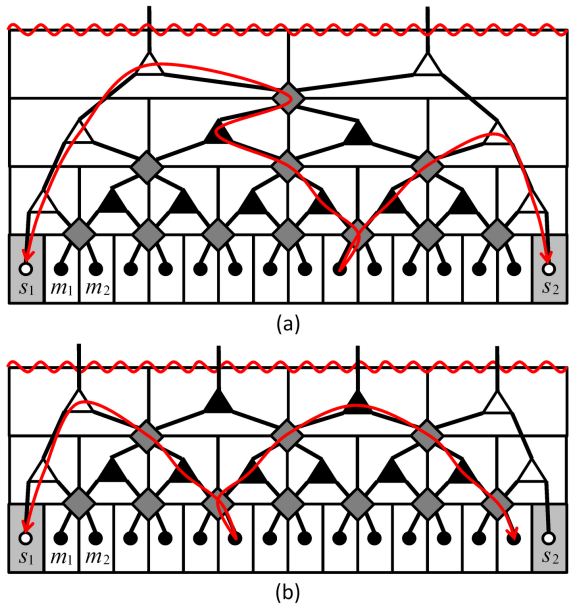}
\end{center}
\caption{Holographic representation of the RKKY interaction: (a) low termerature case, (b) high temperature case. In Fig.(b), two impurities can not interact with each other. Red lines are bending in these figures, but they are geodesic lines.}
\label{fig4}
\end{figure}

The consistency between the main result Eq.~(\ref{kondo}) and the exact $T_{K}$ shows the powerfulness of the MERA analysis. The present method can be easily applied to other systems. For instance, let us consider the competition between the Kondo singlet formation and the Ruderman-Kittel-Kasuya-Yoshida (RKKY) internaction in a two-impurities case~\cite{Doniach}. We assume that the impurities are located at both edges in the 1D chain. Then, the RKKY interaction occurs through the holographic tree. When we consider finite temperature cases, some upper layers of the MERA network are terminated. Then, the distance between two impurities that can interact with each other is restricted. As shown in Fig.~\ref{fig4}(b), two impurities can not contanct with each other at high temperature cases, since the two geodesic curves though the conduction electron is too short to combine these impurities. At the same time, the partial screening of each impurity is originated only from near-neighbor electrons. Even in such a case, the Kondo coupling occurs for large $J$, since $J$ is originally local interaction. On the other hand, at low temperatues and for relatively small $J$ values, the RKKY interaction wins. As shown in Fig.~\ref{fig4}(a), the free energy for RKKY is lower than that for Kondo interaction, since the thermal entropy (or geodesic distance) is larger in the former case. Our view seems to be consistent with the scaling relations $T_{K}\sim e^{-1/J}$ and $T_{RKKY}\sim J^{2}$ for weak $J$.

Summarizing, we have found two types of entanglement in the MERA network of the impurity Kondo model: one comes from finite-temperature effects on the network geometry, and the other represents the screening of the impurity by the conduction electrons. As the latter starts to dominate the electronic state, the Kondo physics emerges. We determine the Kondo temperature $T_{K}$ in a gravitational viewpoint where Eq.~(\ref{key}) plays an important role on connecting the network with the dicrete AdS space. Since $T_{K}$ is essentially an universal parameter, this temperature scale automatically appears from the network geometry itself. Quite recently, it has been discussed that the one-particle Schr$\ddot{\rm o}$dinger equation for the Wilson Hamiltonian is similar to a boundary field theory in the background AdS${}_{3}$ spacetime~\cite{Okunishi}. These recent results illustrate close relationship between the renormalization group approach in condensed matter physics and the AdS/CFT correspondence.

\begin{acknowledgements}
H. M. thanks Tadashi Takayanagi and his coworkers for fruitful discussion beyond ones' expertise. H. M. also thanks T. T. for giving me his manuscript addressing a related topic prior to publication~\cite{Nozaki}.
\end{acknowledgements}

\end{document}